\documentclass[aps,prl,twocolumn,showpacs,superscriptaddress]{revtex4}

\usepackage{graphicx}

\begin{document}

\newcommand{\be}{\begin{equation}}
\newcommand{\ee}{\end{equation}}
\newcommand{\bea}{\begin{eqnarray}}
\newcommand{\eea}{\end{eqnarray}}

\title{Non-equilibrium ribbon model of scroll waves with twist}

\author{Blas Echebarria}
\affiliation{
Departament de F\'{\i}sica Aplicada, Universitat Polit\`ecnica de Catalunya,
Doctor Mara\~n\'on 44, E-08028 Barcelona, Spain.}
\author{Vincent Hakim}
\affiliation{
Laboratoire de Physique Statistique CNRS-UMR8550, Ecole Normale
Sup\'erieure, 24 rue Lhomond 75231 Paris, France.}
\author{Herv\'e Henry}
\affiliation{
Laboratoire de Physique de la Mati\`ere Condens\'ee CNRS-UMR7643, Ecole Polytechnique Palaiseau, France.}
\date{\today}
\begin{abstract}
We formulate a reduced model to analyze the motion of the core
of a twisted scroll wave. The model is first shown to provide a simple description of 
the onset and nonlinear evolution of the
helical state appearing in the sproing bifurcation of scroll waves.
It then serves to examine the experimentally studied case of a medium with spatially-varying
excitability. The model shows the role of sproing in this more
complex setting and highlights the differences between the convective and absolute sproing 
instabilities. 
\end{abstract}
\maketitle

Scroll waves,
spirals tridimensional (3D)
counterparts, are
essential structuring elements
of the dynamics of thick 
excitable media and 
are thought to play an important role in
ventricular fibrillation \cite{Wi98,Wi94}. 
%Rotating waves  are  essential structuring elements
%of the dynamics of 
%excitable media and have been implicated
%in several heart rhythm dysfunctions \cite{Wi98}. 
%Scroll waves,
%spirals tridimensional (3D)
%counterparts, are, in particular, thought to play an important role in
%ventricular fibrillation \cite{Wi94}. 
This has motivated detailed examinations of their instabilities,
both
with chemical reactions in gels \cite{PeAl90,StRo03}
and  
theoretically \cite{He89,Ke88,Bi94,HeHa02,AlSa03,ZaMi04}. Winfree {\it et al.} 
discovered that twist  
can destabilize  a scroll straight core  and lead it to adopt a helical
shape \cite{He89}. This ``sproing'' bifurcation resembles
the twist-induced
instabilities of elastic rods \cite{Love,GoPo98} or of DNA \cite{Cha04}, 
but has remained 
somewhat of a puzzle since,
using long-wave expansions \cite{Ke88},
 the dynamics of a scroll core filament 
was found to be independent of twist \cite{Bi94}. Here, 
insights gained from a numerical stability analysis
\cite{HeHa02} 
%More recently, a numerical stability analysis 
%has shown 
%that sproing appears above a threshold twist and
%a finite wavelength away from the scroll
%wave translation mode \cite{HeHa02}. These results 
lead us
to  formulate 
%in this letter,
a simple model of the core dynamics 
%of the core 
of a twisted scroll wave that
is analytically tractable and agrees semi-quantitatively with the results
of reaction-diffusion (RD) simulations. We first
show that the model provides an easy understanding as well as an accurate
description of sproing. 
Systematic variations of electrophysiological properties are
known to exist in the heart 
and gradients of excitability have been shown to promote
scroll wave instabilities 
%in several experiments 
in chemical media \cite{PeAl90,StRo03}. Therefore, we then choose
the case of a medium with spatially varying  
excitability to test 
the usefulness of the approach
beyond the simplest case of a homogeneous medium.
The results 
show that the observed
instabilities \cite{PeAl90,StRo03} are tightly linked to sproing and
illustrate the subtleties
brought by the problem non-equilibrium setting.

The center of rotation of a planar spiral becomes for a three-dimensional
scroll wave the line of instantaneous
center of rotations $\mathbf{R}(\sigma, t)$ where
$\sigma$ parameterizes this center filament and $t$ is time.
We choose to describe the scroll wave core as the ribbon 
$(\mathbf{R}(\sigma, t), \mathbf{p}(\sigma,t))$, 
 where the unit vectors $\mathbf{p}(\sigma,t)$ point 
orthogonally from the center filament to the line of instantaneous scroll wave
tips.
The local rotation of the line of instantaneous tips 
around the center filament defines the scroll wave twist $\tau_w$,
\begin{equation}
\tau_w=(\mathbf{p}\times\partial_s\mathbf{p})\cdot \mathbf{T},
\end{equation}
where $\mathbf{T}=\partial\mathbf{R}/\partial s$ is the local tangent to
the center filament and $s$ its arclength (with $\partial_s$ simply a notation
for $1/\vert\partial\mathbf{R}/\partial \sigma\vert\ \partial_{\sigma}$).
Starting from a straight, untwisted scroll wave, 
a  gradient
expansion \cite{Ke88,Bi94,HeHa02,GaOt97}
based on the translational and rotational invariance neutral modes 
shows that, at lowest order, the scroll core motion 
is simply driven by  its center filament curvature $\kappa$ 
\cite{Ke88, Bi94,HeHa02} 
\be
{\bf R}_t \cdot {\bf N}=a_1\, \kappa, \;\; {\bf R}_t \cdot {\bf B}=a_2\, \kappa,
\label{ec.mean}
\ee 
where ${\bf N}$ is the filament normal, $\kappa {\bf N}=\partial{\bf T}/\partial s$,
and ${\bf B}={\bf T}\times {\bf N}$ its binormal. 
The case when $a_1 < 0$ is analogous to the filament having  
a negative line tension, and the allied instability
 has been extensively studied 
\cite{Bi94,HeHa02,AlSa03,ZaMi04}. However,
Eq.~(\ref{ec.mean}) leaves twist-induced
instabilities unexplained, since the motion of the
mean filament is not influenced by the ribbon twist \cite{Bi94}. 
Twist appears
at higher orders in the gradient
expansion of ref.~\cite{Ke88,Bi94,HeHa02}
%, reformulated in a slightly different form
%in ref.~\cite{HeHa02}. 
%However, 
but,
besides being somewhat cumbersome, this rigorous approach suffers from
the fundamental difficulty  that sproing sets in at a finite wavenumber
\cite{HeHa02}.
%instability 
Consequently, it
%therefore 
%that it 
cannot be precisely described by a gradient
expansion cut to any finite order
\cite{expansion}.
Therefore, we find it instructive to formulate here a 
simple phenomenological
model
that captures the essence of the phenomenon and that contains
only terms essential for
the instability description. The filament velocity in its normal plane
is written as a generalization of Eq.~(\ref{ec.mean})
\begin{eqnarray}
[\mathbf{R}_t]_{\perp}&=& a_1 \mathbf{R}_{ss}+a_2 \mathbf{R}_{s}\times
\mathbf{R}_{ss}
+ d_1 \tau_w \mathbf{R}_s\times \mathbf{R}_{sss}\label{rdyn.eq}\\
                 &-&d_2 \tau_w [\mathbf{R}_{sss}]_{\perp}
-b_1 \tau_w [\mathbf{R}_{ssss}]_{\perp}
-b_2 \mathbf{R}_s\times \mathbf{R}_{ssss},\nonumber
\end{eqnarray}
where the brackets  
denotes the component of the enclosed vector 
orthogonal to the filament tangent (e.g.
$[\mathbf{R_t}]_{\perp}\equiv\mathbf{R_t -(R_t\cdot T) T}$).
It is worth remarking that the helical instability of an 
elastic ribbon with a gradient dynamics based on extension and
curvature and twist energies \cite{GoPo98} essentially depends on the $a_1, b_1$ and $d_1$
terms.
The $a_2,b_2$ and $d_2$ terms
describe motion in the orthogonal direction. They can
appear in the present non-potential problem due to the handedness of the
spiral rotation and their sign depends on the spiral sense of rotation.
Eq.~(\ref{rdyn.eq}) needs to be completed by the evolution of the ribbon 
twist. The twist kinematics can be adapted
from previous investigations of elastic ribbons.
Following ref.~\cite{KlTa94},
we note that as one slides along the central filament at a fixed time $t$,
the ribbon vector $\mathbf{p}$ rotates and remains orthogonal to the filament
tangent $\mathbf{T}$, 
\begin{equation}
\frac{\partial\mathbf{p}}{\partial \sigma}=
\tau_w \frac{\partial s}{\partial \sigma} \left (\mathbf{T}\times
\mathbf{p}\right )-
\left (\frac{\partial \mathbf{T}}{\partial\sigma}\cdot\mathbf{p}\right ) \mathbf{T}.
\label{twfil}
\end{equation}
This is also true as time evolves when one stands at a fixed abscissa
$\sigma$ and similarly,
\begin{equation}
\frac{\partial\mathbf{p}}{\partial t}=\omega\left (\mathbf{T}\times
\mathbf{p}\right )-
\left (\frac{\partial \mathbf{T}}{\partial t}\cdot\mathbf{p}\right ) \mathbf{T},
\label{twtime}
\end{equation}
where $\omega$ is the local instantaneous scroll wave rotation frequency.
Comparing crossed-derivatives of Eqs.~(\ref{twfil}, \ref{twtime}) gives
as single compatibility condition the equality of the projections
of $\partial_{t,\sigma}\mathbf{p}$ and $\partial_{\sigma,t}\mathbf{p}$
on $\mathbf{T\times p}$,
\begin{equation}
\frac{\partial}{\partial t}\left (\tau_w \frac{\partial s}{\partial
  \sigma}\right )=
\frac{\partial \omega}{\partial \sigma}
+ \left (\frac{\partial \mathbf{T}}{\partial \sigma}\times 
\frac{\partial \mathbf{T}}{\partial t}\right )\cdot
\mathbf{T}.
\label{tweq}
\end{equation}
The kinematic Eq.~(\ref{tweq}) is a local description for an extensible
ribbon of the well-known conversion of
twist into writhe \cite{Ka02} associated to linking number conservation at a
global level. The specific dynamics of
the present problem 
is encoded in 
the twist-dependent rotation frequency $\omega$ . A good approximation
for moderate twist is obtained by keeping the first twist corrections
to the untwisted scroll frequency $\omega_0$,
\begin{equation}
\omega= \omega_0+ c\, \tau_w^2 +D {\partial_s}\tau_w+(\mathbf{T}\cdot\partial_t 
\mathbf{R})\tau_w,
\label{alpdyn}
\end{equation}
where 
the coefficients $D$ and $c$ can be explicitly calculated by linearization
around the straight scroll wave and projection over the adjoint
eigenmodes \cite{HeHa02}. The last term in Eq.~(\ref{alpdyn}) is due
to the apparent rotation of $\mathbf{p}$ coming
from changing position along the filament. Eq.~(\ref{tweq}) with (\ref{alpdyn})
is equivalent to Keener's phase equation \cite{Ke88} and completes our formulation
of the ribbon model. In the following,  this simplified
model is compared to  simulations of RD equations
in the form
\cite{Barpra}
\be
\partial_t u=\nabla^2 u + u(1-u)[u-(v+\beta)/\alpha]/\epsilon,\;\;
\partial_t v=u-v, \label{ec.barkley}
\ee
with $\alpha=0.8$, $\epsilon=0.025$, and different values of $\beta$.
Eqs.~(\ref{ec.barkley}) are simulated with an explicit second order scheme, with
$dx=0.15$, and $dt=5.625\cdot 10^{-3}$.
 
- {\em Sproing.} 
Taking a vertical  filament along the z-axis 
and assuming small transverse
$X, Y$ deformations, 
Eqs. (\ref{rdyn.eq},\ref{tweq},\ref{alpdyn}) become to quadratic order,
\begin{eqnarray}
\partial_t W &=& a \partial^2_z W + id\tau_w \partial^3_z W - b \partial^4_z W, \label{ec.W}\\
\partial_t\tau_w &=& \partial_s (D \partial_s \tau_w) + \partial_s (c\tau_w^2) +\partial_s \omega_0\nonumber \\
&&+\mathrm{Re} \{ [\partial_z(\tau_w \partial_z \overline{W})-i(\partial^2_z
  \overline{W})\partial_z]\partial_t W\},\ \
\label{ec.tauW}
\end{eqnarray}
where a complex notation has been used for the deformation
field 
$W(z,t)=X(z,t)+i Y(z,t)$ 
and the constants $a,b,d$ (e.g. $a=a_1+i a_2$), and $\partial_s$ can be approximated by
$[1-\vert \partial_z W\vert^2/2]\partial_z$.
For a uniformly twisted filament in a homogeneous medium ($\omega_0= \mathrm{cst.}$), the linear modes 
$W(z,t)= e^{ikz+\Omega t}$, correspond to helices of
pitch $k$.
Their dispersion relation 
is obtained from Eq. (\ref{ec.W}) as,
\be
\Omega=-a k^2 + d \tau_w k^3 -b k^4.
\label{ec.disp}
\ee 
\begin{figure}
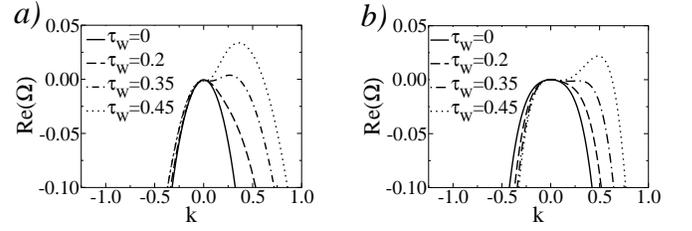

\centering
\includegraphics[width=4.cm]{fig1a.eps}\hspace{0.5cm}
\includegraphics[width=4.cm]{fig1b.eps}\\[0.2cm]
\caption{a) Dispersion relations obtained from a direct linear stability
analysis of a twisted scroll wave of Eq.~(\ref{ec.barkley}) \cite{HeHa02}, with
$\alpha=0.8$,
$\epsilon=0.025$, and $\beta=0.01$, and b) from Eq. (\ref{ec.W}), with
$a=0.2+0.2i$, $d=3.5+i$, and $b=2+i$, chosen to give a
semi-quantitative overall agreement between the two sets of curves. Above a crit
ical
twist, $\mathrm{Re}(\Omega)$ is positive for $k$ around a non-zero $k_c$
and the straight scroll becomes unstable to a finite
wavenumber Hopf instability. Since the instability occurs at finite
wavelength, a better fit at $k=0$ (using the exact value of $a$
\cite{HeHa02}) typically deteriorates the overall fit.
\label{fig.disp}
}
\end{figure}
With appropriate constants $a,b,d$, it is 
%qualitatively 
similar to 
the dispersion relation obtained 
from RD Eq.~(\ref{ec.barkley})
(Figs. \ref{fig.disp}a,b). A secondary local maximum appears 
away from $k=0$ when $\tau_w=4/3 \sqrt{2 a_1 b_1/d^2_1} $. 
In a large box, instability sets in at the critical twist
 $\tau_w^c=2\sqrt{a_1
b_1/d^2_1}$, with the pitch of the allied helix equal to 
$k_c=\sqrt{a_1/b_1}$. The mode $k$ becomes
unstable above $\tau_{w,k}^c$ when $Re[\Omega(k)]>0$. 
For a homogeneous twist $\tau_w$ slightly above 
$\tau_{w,k}^c$,  
the radius $R(t)$ of a helix of pitch $k$  grows as
\be
R_t=\gamma_k (\tau_w-\tau_{w,k}^c) R,
\label{rgl}
\ee
where $\gamma_k=Re[d\Omega(k)/d\tau_w]=d_1 k^3$.
Saturation of the instability comes
from the coupling between twist and bending described by Eq.~(\ref{ec.tauW}).
For an helical mode of pitch $k$ and time dependent but homogeneous twist
and radius, the partial $s$ derivative terms of Eq. (\ref{ec.tauW}) vanish.
The last and only remaining term is equal to $-k^2 (\tau_w+k) R R_t$ so
integration of
Eq. (\ref{ec.tauW}) 
%(i.e. conservation of linking number for a deforming helix)
shows that the twist $\tau_w$ decreases with the helix
radius,  
\be
\tau_w= \tau_w^0- (\tau_w^0+k)k^2 R^2/2, \label{cons.link}
\ee
where $\tau_w^0$ is the initial twist of the straight scroll, and we have
assumed $(Rk)^2\ll 1$ \cite{footlink}.
Comparison of Eq.~(\ref{rgl}) and (\ref{cons.link}) describes sproing as 
 a supercritical bifurcation
\be
R_t=\gamma_k(\tau^0_w-\tau^c_{w,k}) R - \frac{\gamma_k}{2} (\tau_w^0 + k) k^2 R^3.
\ee
The deformation
of the center filament  decreases the initial twist until the critical
value $\tau_w=\tau_{w,k}^c$, is reached at which point the driving force 
for the instability disappears. The final helix radius is
$R=[2 (\tau^0_w-\tau^c_{w,k})/(\tau_w^0 + k) k^2]^{1/2}$ (with $k=k_c$ in a large box).

These analytic results compare well to results of RD simulations 
with periodic boundary
conditions (BC) in the z-direction to enforce linking number
conservation \cite{epaps}.
As previously reported \cite{HeHa02},
sproing is found to be a supercritical bifurcation and the twist of a bifurcated
helix is very close to the critical one, in good agreement with the above
findings. In large systems, as for oscillatory media \cite{RoCh98}, the
helices resulting from sproing may be unstable to secondary Hopf instabilities 
\cite{HeHa02}  which appear sensitive
to higher order nonlinearities not included in Eq.~(\ref{rdyn.eq}).
These can be described by  amplitude equations for the coupled
helix amplitude and excess local linking number. The equations
can be derived from the reduced
model or directly from the RD Eq.~(\ref{ec.barkley}) and take a form
similar to other cases with a conservation law \cite{hr99}.
In simulations of Eq. (\ref{ec.barkley}) these secondary instabilities 
typically result in other helices with smaller wavenumber, or in modulated
structures.
\begin{figure}
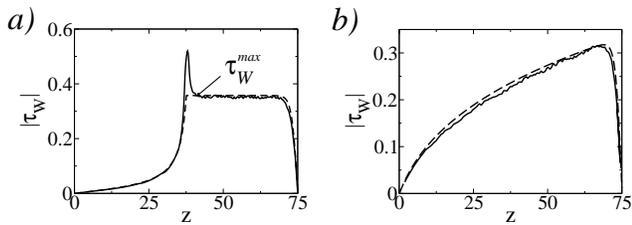

\centering
\includegraphics[width=4.0cm]{fig2a.eps}\hspace{0.2cm}
\includegraphics[width=4.0cm]{fig2b.eps}
\caption{Distributions of twist along a straight filament in  simulations of
RD Eq.~(\ref{ec.barkley}) (solid line), and given by Eq. (\ref{ec.tauW}) (dashed
 line).
a) Jump of excitability obtained  by taking
  $\beta(z)=\beta_b+(\beta_t-\beta_b)\Theta(z-L/2)$ in Eq.~(\ref{ec.barkley})
or in the model by taking the allied spiral frequency jump
  $\omega_0=\omega_b+(\omega_t-\omega_b)\Theta(z-L/2)$.  b)
  Linear gradient of excitability: $\beta(z)=\beta_b+(\beta_t-\beta_b)z/L$ in
Eq.~(\ref{ec.barkley}) or
   $\omega_0=\omega_b+(\omega_t-\omega_b)z/L$ for the model. The
  parameters are the same as in Fig. \ref{fig.disp}, with $\beta_b$=0.01,
  $\beta_t=0.03$, $\omega_b=1.80$, $\omega_t=1.696$, and
  coefficients $D=0.578$, $c=0.720$, at the bottom, and $D=0.614$, and
  $c=0.856$, with equivalent expressions. This gives a theoretical prediction of
 $\tau^{max}_w \simeq
(\Delta \omega_0/c)^{1/2}=0.35$ in a). Here and in the following we plot twist i
n absolute
  value.}
\label{fig.twist}
\end{figure}

\begin{figure}
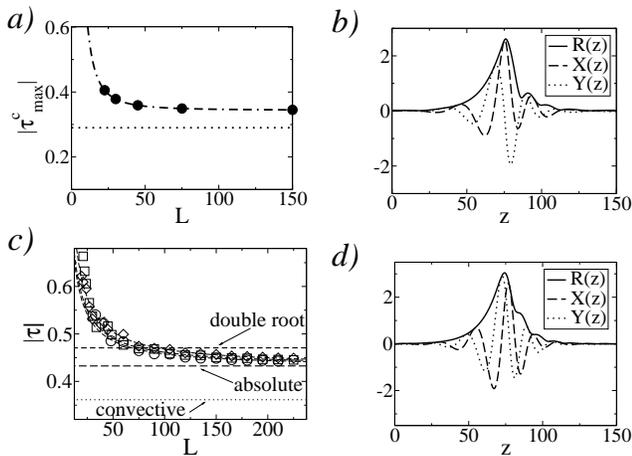

\centering
\includegraphics[width=4cm]{fig4a.eps}\hspace{0.2cm}
\includegraphics[width=4cm]{fig4b.eps}\\
\includegraphics[width=4cm]{fig4c.eps}\hspace{0.2cm}
\includegraphics[width=4cm]{fig4d.eps}
\caption{a) Instability onset {\em vs} domain size
  $L$, for domains with a jump in excitability.  For fixed $L$, $\beta_t$
  is increased until the system becomes unstable. The corresponding
critical twist (filled circles) is plotted together with the fit (dashed line):
  $\tau^c_{max}=0.344+31.3/L^2$. For comparison, the
  onset of sproing for periodic BC ($\tau^c_w \simeq 0.29$) is also shown (dotte
d line)
  for $\beta=0.03$, that corresponds to the critical value of $\beta_t$ in a lar
ge
  system.  b) Solutions for
  $\beta_b=0.01$ and $\beta_t=0.031$.  c) Model
 critical twist
  obtained by solving the eigenvalue problem {\em vs} $L$ for the distribution
 of twist
  shown in Fig. \ref{fig.twist}a (diamonds),
a constant twist in half of the system
  (squares), and
{\em vs} $L/2$ for a constant twist in the whole system (circles).
  For comparison we
  also show the onset of convective (dotted line), absolute
  instability (long dashed line), and the onset obtained with the double root
  criterium (short dashed
  line). d) Critical modes obtained from the eigenvalue problem, using the
  distribution of twist, for $L=150$.}
\label{fig.abs}
\end{figure}

- {\em Inhomogeneous twist.} 
Most experimental 
situations correspond to imposing
free non-flux BC  on Eq.~(\ref{ec.barkley}) 
rather than periodic ones.
These do not conserve total linking number and an initially twisted scroll
wave 
untwists \cite{PeAl90} in an
homogeneous medium. Spatial variations of excitability do however promote twist
formation.
Fig. \ref{fig.twist} shows RD simulations for a straight vertical filament, with
 non-flux BC, 
 and either a jump (Fig. \ref{fig.twist}a), or a linear gradient
 (Fig. \ref{fig.twist}b) of the value of $\beta$ in the z-direction.  
For moderate variations of $\beta$, the 
initial  
untwisted core, remains straight but
evolves toward a final twisted and steadily rotating configuration.
The different natural spiral rotation
frequencies $\omega_0 (z)$ create phase differences between different heights
$z$ which together with
the rotation frequency increase with twist [Eq.~(\ref{alpdyn})]
lead to a steady state.
The resulting
distribution of twist 
can be computed,
either analytically or numerically, from the model Eq.~(\ref{ec.tauW}) using
the appropriate source term $\partial_z\omega_0$.
The calculated distribution of twist agrees
remarkably well with the RD simulations as shown in Fig. \ref{fig.twist}.
\begin{figure}
\centering
\includegraphics[width=4cm]{fig3a.eps}\hspace{0.cm}
\includegraphics[width=4cm]{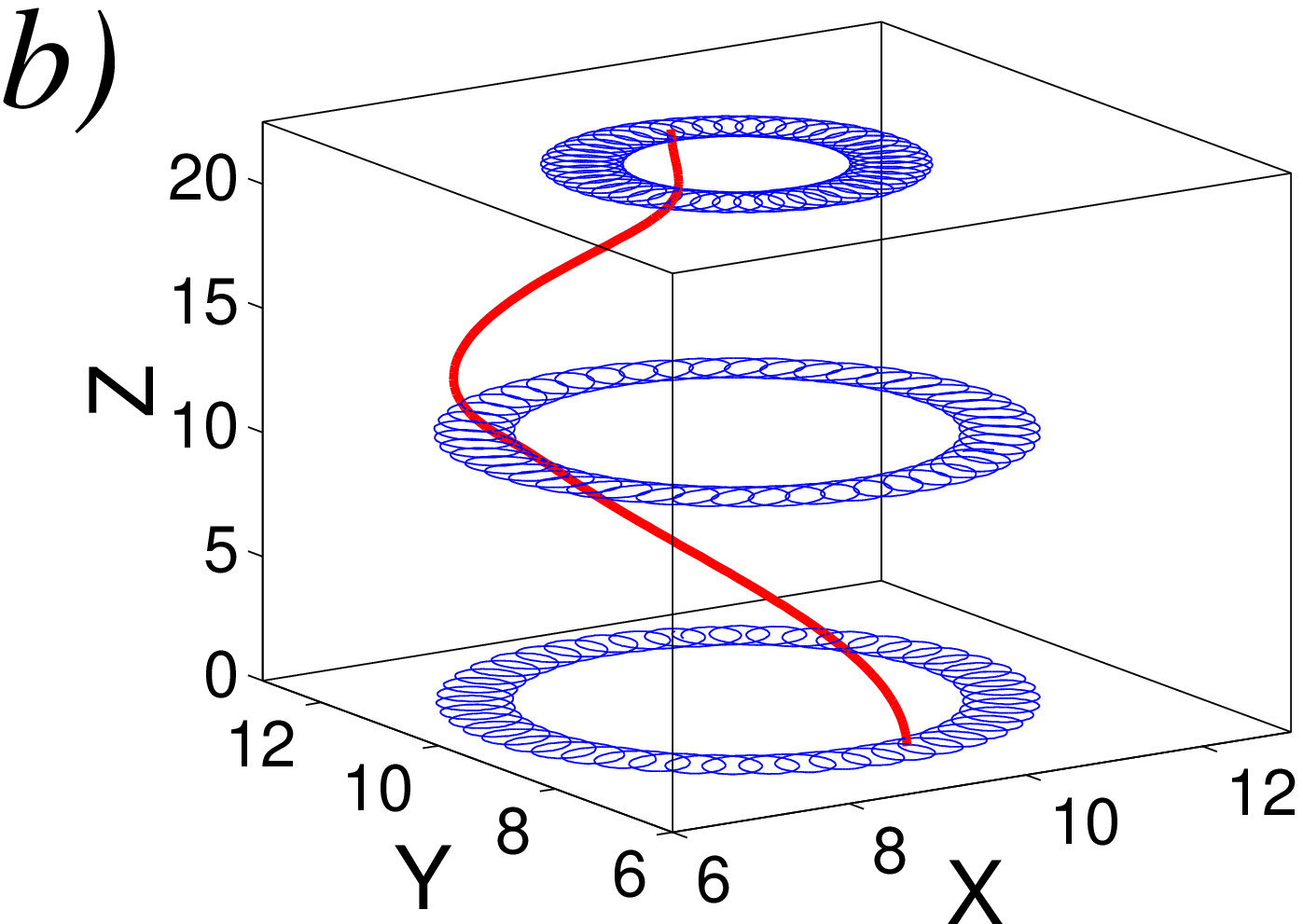}
\caption{a) Center filament maximal radius {\em vs}
  $\beta_t$. The straight scroll becomes unstable at
  $\beta_t^c \simeq 0.03939$, resulting in a small hysteresis. b)
A 3D view of the solution for $\beta_t=0.0399$.}
\label{fig.inst}
\end{figure}
For definiteness, we focus on the case of a medium
with an excitability jump.
In the RD Eqs. (\ref{ec.barkley}), we fix $\beta_b=0.01$ in the medium
bottom half
part, and take $\beta=\beta_t>\beta_b$ in the less excitable top half.
When the jump in excitability is larger than a critical value,
the straight scroll becomes unstable, very similarly
to what is observed in experiments \cite{PeAl90,StRo03,foot3}.
The instability is slightly subcritical and
the resulting structures modulated helices (Fig. \ref{fig.inst}). 
To clearly relate this instability to
sproing,  we consider now the limit of large
systems. Then, for a moderate jump of excitability, the scroll core
is straight and its frequency is
basically set by the domain most excitable half where the scroll twist 
is negligible.
The scroll
twist $\tau^{max}$
in the domain less excitable half is almost constant  and simply determined
by the frequency jump $\Delta \omega_0$ between the two domain parts,
$\tau^{max}_w \simeq (\Delta \omega_0/c)^{1/2}$.
When $\tau_w^{max}$ reaches the
sproing threshold, 
for a large enough jump, 
one could expect sproing to set in
with the center filament taking the shape of a helix
of constant radius in the low excitability region and radius decreasing
to zero in the higher excitability part. 
However, the instability onset differs from the sproing threshold in a homogeneous system, even
when $L \rightarrow \infty$ (Fig. \ref{fig.abs}a). Furthermore,
the  bifurcated filament radius decreases  exponentially
also in the region of constant
twist (Fig. \ref{fig.abs}b).
In  order to clarify the phenomenon, we have analyzed the ribbon model 
in this geometry.
We have
solved the eigenvalue problem given by Eq. (\ref{ec.W}), with the distribution of
twist calculated with Eq. (\ref{ec.tauW}) (with constant values of
$D=0.578$ and $c=0.72$). Similarly to RD simulations, an instability
develops
 in the region of constant twist when $\tau_w^{max}$ is large enough
but its threshold 
differs from the sproing threshold for  
periodic domains
(Fig. \ref{fig.abs}c).
The instability is nonetheless
related to sproing. The reason is that periodic BC allow the
growth of convective instabilities, that decay with non-flux
BC. For a complex growth
rate $\Omega$,  four complex wavenumbers $k_i(\Omega)$ satisfy
the dispersion relation Eq.~(\ref{ec.disp}). The relevant sproing
absolute spectrum, for a given constant twist in a large domain,
lies on the curve of complex
$\Omega$ such that $Im[k_2(\Omega)]=Im[k_3(\Omega)]$, with the $k_i$ 
ordered by increasing
imaginary part \cite{Ku,SaSc00}. For low twist, 
this curve lies entirely in the $Re(\Omega)<0$
half plane. 
 The  absolute instability threshold  
twist, for which the curve
crosses the $\Omega$ imaginary axis \cite{ToPr98},
coincides with the large $L$ limit of 
$\tau_c(L)$ as shown in Fig. \ref{fig.abs}c.
The most unstable modes at threshold are two counterpropagating waves, with
the same spatial growth rate, and nonzero group velocity. The 
similarity 
between the critical modes for the ribbon model and the RD equation 
(Fig. \ref{fig.abs}d) further shows that  
sproing is also a likely explanation of the latter case and of the experimental
 observations \cite{PeAl90,StRo03}.

In conclusion,
the proposed ribbon model provides 
a semi-quantitative description
of the motion of twisted scroll waves and a clear understanding of several
features that are difficult to directly extract from RD equations. 
This will hopefully help to further analyze scroll wave dynamics in complex media 
and to better assess   
the effects of gradients of 
electrophysiological properties and other complicating features 
in the cardiac muscle.

B.E. acknowledges financial support by MCyT (Spain), and by MEC
(Spain), under project FIS2004-02570.

\newpage

\vspace{2cm}

\setcounter{figure}{0}

\begin{figure}
\centering
\includegraphics[width=4.5cm]{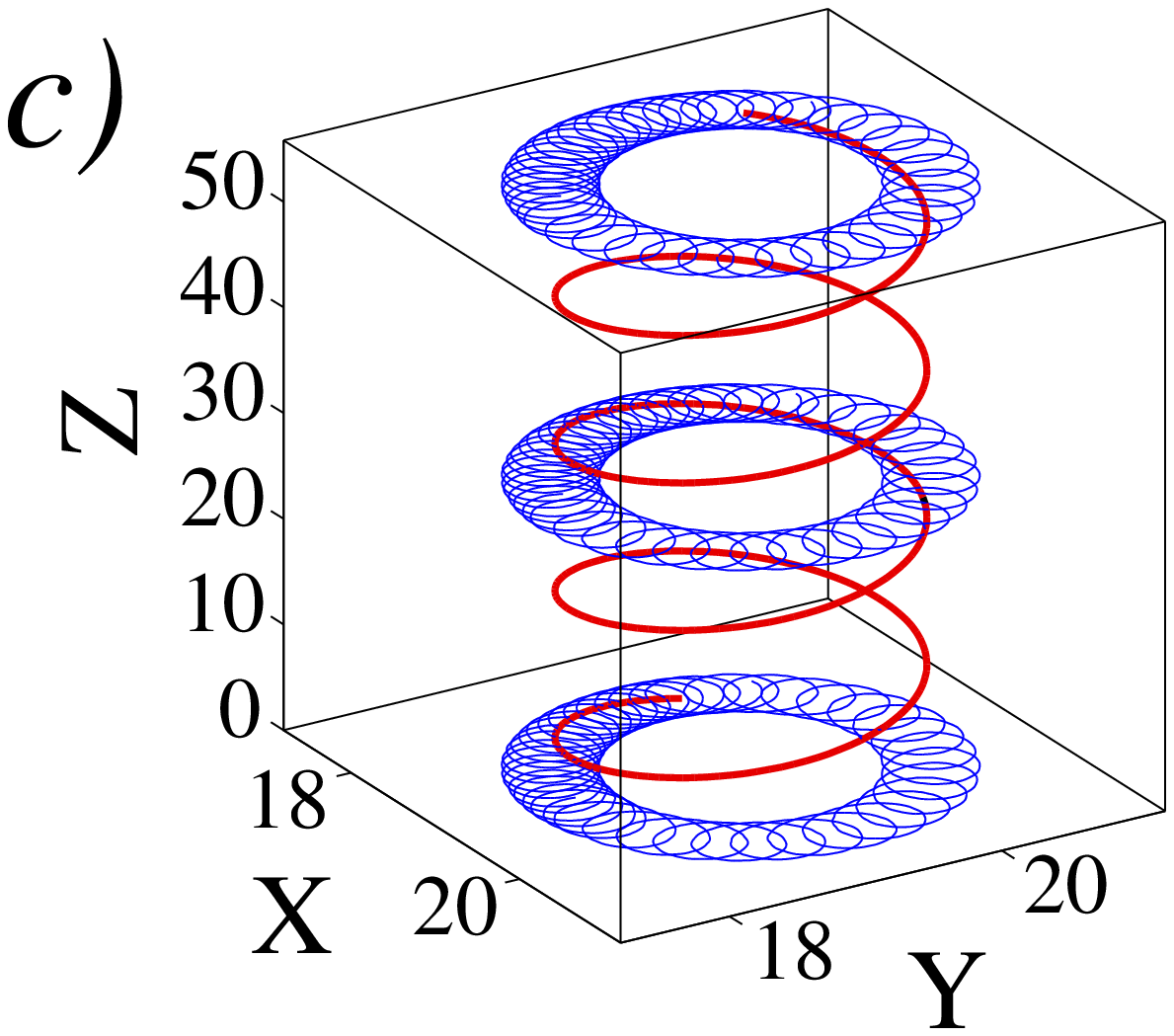}\hspace{-0.2cm}
\includegraphics[width=4cm]{fig1d.eps}
\caption{
%a) Dispersion relation obtained from a direct linear stability
%analysis of a twisted scroll wave of Eq.~(\ref{ec.barkley}) \cite{HeHa02}, with
%$\alpha=0.8$,
%$\epsilon=0.025$, and $\beta=0.01$, and b) from Eq. (\ref{ec.W}), with
%$a=0.2+0.2i$, $d=3.5+i$, and $b=2+i$. Above a critical
%twist, the straight scroll becomes unstable to a finite
%wavenumber Hopf instability. 
Simulations of reaction-diffusion Eq.~(8) of the main text.
The
core of the initial twisted scroll
is 
straight and vertical, and
periodic boundary
conditions are applied in the z-direction to enforce linking number
conservation.
c) Restabilized scroll mean filament, for the
parameters corresponding to Fig.~1 a) of the main text ($\alpha=0.8$,
$\epsilon=0.025$, and $\beta=0.01$), and $\tau_w=0.45$. d) Nondimensional radius of the helix as a 
function of the reduced distance
to the onset of sproing, for several values of $\beta$ (full symbols)
in a small box of height $L$ with a single
turn of helix ($\tau^0_w=k=2\pi/L$). The predicted
straight line of slope one (dotted) is shown for comparison. 
Note that this linear relation holds beyond small departures from threshold
since linking number conservation and restabilization of the
bifurcated helix at the critical twist give 
$\tau_{w,k}^c=\tau_w^0/[1+(R\tau^0_w)^2]$ without any expansion,
as follows from  footnote [19] of the main text with $k=\tau_w^0$. In the
limit $(R\tau_w^0)^2 \ll 1$ this expression reduces to the one obtained from
Eq.~(13).
\label{fig.dispaux}
}
\end{figure}


\begin{thebibliography}{plain}
\bibitem{footlink}
Eq.~(\ref{cons.link}) can be directly obtained from linking number conservation.
The initial straight filament has zero writhe and total twist ${\cal
  T}^i_w=L\tau_w^0/(2\pi)$. The final helix  has a writhe
${\cal W}_r=k\, L \{1-[1+(Rk)^2]^{-1/2}\}/(2\pi)$ and a total length
$L [1+(Rk)^2]^{1/2}$. So its local twist is
$\tau_w=2\pi({\cal T}^i_w-{\cal W}_r) /\{L [1+(Rk)^2]^{1/2}\}$. Expansion for
$(Rk)^2\ll 1$  gives back Eq.~(\ref{cons.link}) and shows that both  
the filament length increase and the creation of writhe contribute to decrease
the local twist.
\bibitem{foot3} The  scroll wave core was observed to break in
Ref.~\cite{PeAl90,StRo03}, presumably due to collisions with the
lateral boundaries.
\end{thebibliography}
\end{document}